# ASSESSING THE MATURITY OF CYBERSECURITY EDUCATION IN VIRGINIA AND IMPACT OF STATE LEVEL INVESTMENT

**Patrick Mero**
Christopher Newport University
e-mail address

**Aaron Pepsin**
Christopher Newport University
e-mail address

**Chris Kreider**
Christopher Newport University
chris.kreider@cnu.edu

**ABSTRACT**

With a global shortage of cybersecurity students with the education and experience necessary to fill more than 3 million jobs, cybersecurity education is an international problem. Significant research within this field has explored this problem in depth, identifying a variety of shortcomings in the cybersecurity educational pipeline including lack of certifications, security clearances, and appropriate educational opportunities within institutions of higher education. Additional research has built on this, exploring specific gaps within what cybersecurity opportunities are provided within institutions of higher education. We build an ordinal scale for assessing this, the cybersecurity education maturity model scale (CEMMs), and provide evidence of reliability and validity. We then calculate the CEMMs score for all public four-year universities in the state of Virginia between 2017 and 2025, with 2017 marking a year in which the state started the Commonwealth Cyber Initiative (CCI). We find that the scale proposed provides a consistent and reliable way to compare the cybersecurity offerings available between universities. When comparing year to year average CEMMs score, we find that public four year universities in Virginia are increasing their program offerings in the area of cybersecurity, with potential to make an impact on the cybersecurity jobs gap.

**INTRODUCTION**

Cybersecurity is an international problem, with an estimated global shortage of over 3 million cybersecurity professionals in 2022, with this number growing significantly between 2015 and 2021 (Towhidi and Pridmore 2023). An analysis of cybersecurity jobs in various subfields found that higher education degrees were frequently required, and expected of job candidates (Ramezan 2023). Some states in the United States of America, such as the state of Virginia, have made multi-million dollar investments to promote research, innovation and talent development through initiatives such as the Commonwealth Cyber Initiative (CCI) (Murphy et al. 2023). Working groups, comprised of faculty and industry leaders across the state convened to develop overarching strategy on how to best achieve these goals, specifically identifying three dimensions of the Cybersecurity Workforce Gap including ensuring universities had appropriate program offerings, and program capacity with a sufficient pipeline of incoming students, highlighted below in figure 1 (Kreider and Almalag 2019).

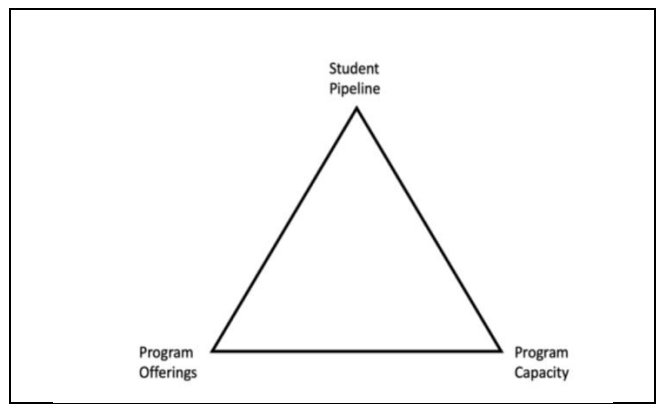

**Figure 1. The Three Dimensions of Meeting Cybersecurity Jobs Needs in Higher Education**

Follow-up work to this proposed a scale assessing cybersecurity related program offerings in institutions of higher education. This scale provided an ordinal assessment of the availability cybersecurity related content, from having no classes dedicated to the topic (a score of 0), through to offering multiple doctoral specializations within the field (a score of 10). In between these scores, a variety of different milestones that reflect academic offerings were categorized, with the scale highlighted below in Table 1 (Kreider and Almalag 2020).

| 0 | 1 | 2 | 3 | 4 | 5 | 6 | 7 | 8 | 9 | 10 |
|---|---|---|---|---|---|---|---|---|---|---|
| No offerings | Class(es) as part of an elective | Associates degree | Minor or area of emphasis | Single Bachelors Degree/ Majors | Multiple Bachelors Degrees/ Majors | Single Masters Degree | Multiple Masters Degrees | Single Doctoral Degree | Multiple Doctoral Degrees | Postdoc Options |

**Table 1. Initially Proposed Maturity Mode Scale of Higher Education Cybersecurity Offerings**

This proposed scale was referred to as a Maturity Model, as higher values in the scale not only represented greater progress towards being capable of providing students with an expected higher education degree, but also capability of contributing qualified faculty capable of increasing the program capacity. While the scale was designed to for evaluating cybersecurity program offerings across a set of universities to assess the impact of state level investments, data was never gathered or presented utilizing the proposed scale.

This work extends the proposed scale by gathering data from public four year universities in the state of Virginia, ensuring the scale adequately captures the various levels of, and assessing change in program offerings from the inception of the Commonwealth Cyber Initiative in 2017 through the 2023-2024 academic year. This is completed via both manual and automatic analysis of publicly available academic catalogs. The rest of this paper will be structured as follows. We will first briefly explore the existing literature on cybersecurity education, specifically the role of higher education in meeting the needs and expectations of the global work force. We will then describe our data collection methodology and present our results. Finally, we will discuss implications of our work, and provide a summary of our findings.

**LITERATURE REVIEW**

Significant work has been generated over the preceding few years exploring the need to fill cybersecurity related positions, and the lack of candidates meeting the necessary qualifications.

Ramezan (2023) generalized the most common themes found in what employers seek in potential applicants based on their applications and analyzed the importance of each expectation. One of the most significant findings is the broad set of skills necessary for Cybersecurity positions, ranging from higher education (degrees) to various certifications (that differ based on sub-field) and security clearances. This highlights the field's complexity, identifying the importance of the multiple facets of cybersecurity employment. This can be seen as a call for higher education institutions to simply do a better job at preparing students for cybersecurity positions, including providing opportunities for certifications and security clearances, as well as better preparing students for the large skill-set and experiences that employers are looking for. Echoing a similar sentiment, Murphy et al. (2023) explores the workforce gap in cybersecurity from three perspectives: Employers, higher education institutions, and students. From a program offering perspective, it points out that educational institutions need to ensure they are staying on top of trending topics to include in their curricula. They argue this falls on the institutions themselves to ensure that they are staying in the loop and doing their best to ensure their students are ready to enter the workforce

From the higher education perspective, Kreider and Almalag (2019) provide insight into the findings of a working group exploring Educational Issues in Cybersecurity (EICS). This group was comprised of 17 members including instructional faculty, directors, deans, provosts and a representative from a state education council that reported back to an overarching advisory council. Utilizing a variety of provided resources related to state level cybersecurity needs including regional cybersecurity workforce needs; cybersecurity related educational options provided by the state; cybersecurity resources available through the National Initiative for Cybersecurity Education (NICE); and details on multiple councils, alliances and partnerships within the cybersecurity domain. These resources were used to develop a gap analysis, identifying three key dimensions that needed to be addressed appropriately to maximize the state level educational contribution towards reducing the jobs gap including program offerings, program capacity and student pipeline, each defined below in table 2.

| Dimension | Narrative |
|---|---|
| Program Offerings | Assuming an infinite supply of students, and infinite resources with which to educate those students, are we capable of offering a set of programs that are sufficient to close the workforce gap in cybersecurity? |
| Program Capacity | Assuming an infinite supply of students, and a set of educational comprehensive programs and offerings, would our educational institutions have the necessary resources to close the workforce gap in cybersecurity? |
| Student Pipeline | Assuming a set of comprehensive educational programs, and infinite resources with which educational institutions could use to implement, would there be enough students willing and capable to engage in such programs? |

**Table 2. Summary of the Gap Analysis Dimensions From (Kreider & Almalag, 2019)**

Additional research has explored how government level funding has played a role in closing this gap. Balon and Baggili (2023) explored how funding from government and other 3rd party agencies have impacted the skill gaps in cybersecurity education. This funding provided opportunities to help motivate and prepare students for the workforce, through initiatives focused on both students and employers. One such example included $62 million dollars funded by the US government in 2013 to assist with the development of the cybersecurity workforce. This funding helped higher-education institutions invest in new ways to develop their curricula, and add ways to enhance the hands on and experiential elements of their cybersecurity programs.

Despite these works, there is little, if any research that quantitatively explores the impacts of these initiatives. While many papers cite an increasing number of open jobs, few if any provide any meaningful measures to assess whether progress is being made towards closing this gap. One such paper by Kreider and Almalag (2020) provides an early attempt at assessing the progress of institutions of higher education towards assisting with closing this jobs gap. The proposed ordinal scale provides a university a score between 0 and 10, depending on what cybersecurity offerings are available within the university. This scale not only measures what cybersecurity options are available for students to contribute back to the workforce in terms of classes, specializations, minors, undergraduate and graduate degrees, but also the overall maturity of the cybersecurity educational ecosystem. The authors refer to this as a maturity model, noting that higher scores not only contribute back to the jobs gap, but eventually provide additional qualified faculty capable of increasing overall program capacities through the granting of masters and doctoral degrees.

**METHODOLOGY**

This project proceeded, starting with a manual assessment of program maturity using the maturity scale proposed by Kreider and Almalag (2020), with the intent of either confirming it adequately captured the identified levels of maturity, or modifying it to be more robust. Using the manually gathered data, the scale was modified, and prior assessments were re-evaluated in the context of the new scale. Finally, an automated tool for analyzing text frequencies and course offerings in university catalogs was used to provide evidence of reliability and validity of the scores assigned to various state universities.

**Updated Scale**

During manual empirical analysis of academic catalogs utilizing the original scale proposed by Kreider and Almalag (2020), a few of the ordinal levels were identified as irrelevant, while others were missing. In the bigger picture, their scale focused started with associates degrees at 1, and ended with post doctoral options. Notably, their scale missed opportunities to categorize cybersecurity specializations, certificates, concentrations or modules at the graduate level. Upon assessment of the graduate cybersecurity offerings in the state of Virginia, much is provided as a subset of another degree, such as computer science or information technology. The updated scale is reflected below in table 2.

| 0 | 1 | 2 | 3 | 4 | 5 | 6 | 7 | 8 | 9 | 10 |
|---|---|---|---|---|---|---|---|---|---|---|
| No offerings | Class(es) as part of an elective | Minor or area of emphasis or concentration | Single Bachelors Degree/ Majors | Multiple Bachelors Degrees/ Majors | Masters Area of Emphasis/ Specialization, concentration or module, or certificate | Single Masters Degree | Multiple Masters Degrees | Doctoral Degree Area of Emphasis/ Specialization | Single Doctoral Degree | Multiple Doctoral Degrees |

**Table 3. Updated Cybersecurity Education Maturity Scale**

**Manual Catalog Assessment**

All public four year universities in the state of Virginia were evaluated from years 2016-2024. Initially, all catalogs were manually evaluated by a researcher. This process started with an evaluation of the current catalog year using any available search resources to identify any cybersecurity related content in the catalog. This primarily included search for broad terms such as security, cyber and assurance, and included additional searches for terms located in the NICE framework (Newhouse et al. 2017). Finally, a manual review of the listed offerings, usually provided at the start of the catalog was performed to ensure that the search terms did not miss anything. From hear, the terms used to locate the relevant programs were identified for each university, and manual search progressed backwards from the current catalog year. For example, if a program in the current catalog year was identified under the title *information assurance*, the prior catalog year was searched for the same term as well. Differences in course offerings, majors, minors and concentrations that were detected between years were explored more thoroughly to understand whether the difference was a true positive or a false positive

**Automatic Data Collection for Validation**

To provide additional evidence of reliability and validity, data was automatically gathered from publicly available university catalogs and catalog archives. The universities we selected were all located in Virginia and have entries from 2017 to 2024-2025 aligning with the timeline of Virginia's increased state-level investment in cybersecurity initiatives. The catalogs used also represent a wide range of research activity levels as classified by the Carnegie Classification of Institutions of Higher Education (R1, R2, and R3). The final list includes Christopher Newport University (CNU), James Madison University (JMU), Virginia Commonwealth University (VCU), William & Mary (W&M), Old Dominion University (ODU), Virginia Military Institute (VMI), Virginia State University (VSU), Norfolk State University (NSU), and Virginia Tech (VT).

Using these pdfs, we developed a program to analyze university catalogs. The program was written in Python and utilizes the PyMuPDf library to extract PDF text content and convert it into analyzable formats. Programming it with this ensured flexibility to accommodate the varying catalog formats without resorting to hardcoding solutions because of the standard format. At its core, the program was made to get the numbers of how often a given word appears, that is filtered through an excluded words list to remove conjunctions, prepositions, articles, or common unrelated words. However, the main functionality of the program in our case is to identify the total occurrences of specific keywords, (e.g., "cybersecurity," "security") as they are the most heavily weighted words in our metric for measuring the growth of cybersecurity programs in schools. The frequency of such terms serves as a proxy for gauging program maturity. By comparing keyword occurrences over time and aligning these with the maturity scale, we can discover the degree to which state-level funding has contributed to program development.

Weighting the words also helps the program pull out every reference of it with the 100 words that surround it, 50 on either side, for the purpose of gaining context as to why it was being used and how it aligns with the scale. Additionally, the program isolates and retrieves detailed course descriptions for cybersecurity-related courses using predefined keywords and recognizing common course code patterns. All outputs, including word frequencies, contexts, and course descriptions, are organized into standardized CSV files. This format allows for seamless integration with tools like Excel, ensuring straightforward comparison and visualization across universities and timeframes. A final program was created that worked with the total CSV files, ones that included the total counts for the heavily weighted words previously mentioned, cybersecurity and security, for all universities across all the catalogs available in our range of years. The program compiles these total files into one final total file that can be used for the analysis of this paper.

**RESULTS**

**Manual Analysis Results**

The results of the manual analysis are presented below in table 4, with 130 academic catalogs being examined. Of the universities assessed, only 5 catalog years were not able to be located via public searches on the web.

| Univ. | Name | 16/17 | 17/18 | 18/19 | 19/20 | 20/21 | 21/22 | 22/23 | 23/24 | 24/25 |
|---|---|---|---|---|---|---|---|---|---|---|
| CNU | Christopher Newport University | 1 | 1 | 1 | 2 | 3 | 3 | 3 | 3 | 3 |
| GMU | George Mason University | 8 | 8 | 8 | 8 | 8 | 8 | 8 | 8 | 8 |
| JMU | James Madison University | 5 | x | 5 | 5 | 5 | 5 | 5 | 5 | 5 |
| LW | Longwood University | 2 | 2 | 2 | 2 | 2 | 2 | 2 | 2 | 2 |
| NSU | Norfolk State University | X | X | X | X | X | 6 | 6 | 6 | 6 |
| ODU | Old Dominion University | 5 | 5 | 6 | 8 | 8 | 8 | 8 | 8 | 8 |
| RU | Radford University | 1 | 2 | 2 | 3 | 3 | 3 | 3 | 3 | 3 |
| UMW | University of Mary Washington | 5 | 2 | 2 | 3 | 3 | 3 | 3 | 3 | 3 |
| UVA | University of Virginia | 1 | 1 | 1 | 1 | 2 | 2 | 2 | 8 | 8 |
| UVAW | University of Virginia at Wise | 1 | 1 | 1 | 1 | 2 | 2 | 2 | 2 | 2 |
| VCU | Virginia Commonwealth University | 6 | 6 | 6 | 6 | 6 | 6 | 6 | 6 | 6 |
| VMI | Virginia Military Institute | 1 | 2 | 2 | 2 | 2 | 2 | 2 | 2 | 2 |
| VSU | Virginia State University | 2 | 2 | 2 | 2 | 2 | 2 | 2 | 2 | 2 |
| VT | Virginia Tech | 5 | 5 | 5 | 5 | 5 | 5 | 5 | 5 | 5 |
| W&M | William and Mary | 1 | 1 | 1 | 1 | 1 | 1 | 1 | 1 | 5 |
| | **Average** | 3.14 | 2.92 | 3.14 | 3.50 | 3.71 | 3.87 | 3.87 | 4.27 | 4.53 |

**Table 4. Cybersecurity Education Maturity Scores for 4 Year Universities in Virginia**

**Automatic Validation Results**

To provide additional evidence of reliability and validity for the manually calculated scores, an automatic tool generated word frequencies for terms related to cybersecurity. Not all academic catalogs were capable of being processed by the tool, as some universities used interactive online catalogs, as opposed to downloadable pdf formats. Of the 72 catalogs that were manually inspected, a total of 52 catalogs were able for automatic processing. The total counts were then averaged for each academic year, with a summary of these averages provided in table 5 below.

| Year | 17-18 | 18-19 | 19-20 | 20-21 | 21-22 | 22-23 | 23-24 | 24-25 |
|---|---|---|---|---|---|---|---|---|
| **Average Occurrence of "Cybersecurity"** | 18.33 | 28.67 | 37.57 | 50.83 | 55.57 | 58.71 | 82.14 | 123.50 |

**Table 5. Average Occurrence of "Cybersecurity" Term in Academic Catalogs**

We find that there has been marked and steady growth in cybersecurity related content in academic catalogs from the 2017-2018 academic year through to the 2024-2025 academic year. These averages were then compared to the average score for each academic year calculated using the cybersecurity education maturity model scale (CEMMs) using correlational analysis. This was completed by comparing the word counts for each academic year, with the CEMMs score for that academic year, for each university. A total of 47 university, catalog year, and measure pairs were available for correlational analysis. Results of the pearson correlation indicated that the positive relationship between CEMMs score, and occurrence of "cybersecurity" in the academic catalog was significant with (r(47) = .609, p<.001) and a r² 0.3709. Additionally, as the CEMMs contains ordinal ranks, spearman rho was calculated which also showed a significant large positive relationship (r(47) = .668, p<.001).

**DISCUSSION**

The initial scale proposed by Kreider and Almalag (2020) to ordinally measure the *program offering* dimension of the jobs gap served as a good foundation, but required minor adjustments during data collection. These changes allowed the scale to more accurately categorize whether cybersecurity offerings (courses, minors, majors, etc) were available, and at what level they were available (e.g. bachelors, masters, doctoral). This updated scale is identified as the *cybersecurity education maturity model scale*, or CEMMs. Using the average CEMMs score calculated across all public for year universities in the state of Virginia, we find that the availability of cybersecurity academic program offerings grew significantly between the years 2017-2018, and 2024-2025. This was represented by a CEMMs score of 3.14 in 2017-1028 to 4.53 in 2023-2024. While these score values were manually gathered by researchers, using automated tools that searched for occurrences of "cybersecurity" in pdf accessible academic catalogs was used to provide additional evidence of reliability, finding that the scores correlated significantly with an increase in the term "cybersecurity" in the academic catalog.

**CONCLUSION**

The jobs gap in cybersecurity can be explored through the three dimensions of program offering, program capacity and student pipeline. This paper proposes an ordinal scale for measuring academic *program offerings* in four-year universities, the *cybersecurity education maturity model scale* (CEMMs). Using the CEMMs scale, data was gathered from all public four-year universities in the state of Virginia to assign a CEMMs score, which was then averaged across the state. Additional data was gathered to provide evidence of reliability through automatic analysis of the pdf available academic catalogs, finding that the assigned scores correlated significantly with the automatically gathered values.

We found that between 2017 and 2025, the availability of academic offerings related to cybersecurity increased significantly, moving from a CEMMs score of 3.14 to 4.53. This period of growth coincides with the operating periods of the Commonwealth Cyber Initiative (CCI) and increased state level funding in cybersecurity education. While this cannot be interpreted as a causal relationship, it does provide evidence that overall, academic institutions are answering the call to provide more graduates in needed areas, specifically in this case, cybersecurity.

This study has several limitations, primarily limited to length restrictions in this paper. First, data was only gathered from the beginning of 2017-2018 academic years, the years when the CCI initiative was started. Additional research could explores the years immediately preceding this period to determine whether there was an increase in velocity of growth in cybersecurity academic offerings or not. Second, the evidence of reliability used a relatively simple automatic analysis of word frequencies in academic catalogs. This is based on the assumption that universities would want their new offerings to be found by current and prospective students. This would require the programs and courses to exist in the catalog and be appropriately named so that they could be found via search. However, as there is no universal naming system for academic offerings, it is possible that the automatic word calculation missed some university offerings with differently or unusually named offerings.

Future research should continue to explore this by exploring a wider range of academic years. Additionally, identifying different states with varying levels of investment in cybersecurity education would allow for a more comprehensive picture, including a better understanding of how investments in cybersecurity education have impacted what universities are offering through use of CEMMs scores.